\documentclass[aps,prd,twocolumn,showpacs,10pt,superscriptaddress,preprintnumbers]{revtex4-1}
\usepackage{amsmath}
\usepackage{graphicx}
\usepackage{amssymb}
\usepackage[colorlinks,citecolor=red]{hyperref}
\usepackage{slashed}

\usepackage{color}    
\usepackage{ulem}

\newcommand{\beq}{\begin{equation}}
\newcommand{\eeq}{\end{equation}}
\newcommand{\bea}{\begin{eqnarray}}
\newcommand{\eea}{\end{eqnarray}}
\newcommand{\nn}{\nonumber}
\def\kslash{\slashed{p}}

\begin{document}

\preprint{
{\vbox {
\hbox{\bf LA-UR-21-24074}
}}}
\vspace*{0.2cm}

\title{Probing the electroweak symmetry breaking with Higgs production at the LHC}

\author{Ke-Pan Xie}
\email{kpxie@snu.ac.kr}
\affiliation{Center for Theoretical Physics, Department of Physics and Astronomy, Seoul National University, 1 Gwanak-ro, Gwanak-gu, Seoul 08826, Korea}

\author{Bin Yan}
\email{Corresponding author: binyan@lanl.gov}
\affiliation{Theoretical Division, Group T-2, MS B283, Los Alamos National Laboratory, P.O. Box 1663, Los Alamos, NM 87545, USA}

\begin{abstract}
The electroweak symmetry breaking (EWSB) mechanism is still an undecided question in particle physics. We propose to utilize the single top quark and Higgs associated production ($th$), $Zh$ production via gluon fusion at the LHC to probe the couplings between the Higgs and the gauge bosons and further to test the  EWSB.  We demonstrate that the $th$ and $gg\to Zh$ productions are sensitive to the relative sign of couplings ($ht\bar{t}$, $hWW$) and ($ht\bar{t}$, $hZZ$), respectively. We find that the relative sign between $hWW$ and $hZZ$ couplings could be fully determined after combining the present measurements from $gg\to h$, $t\bar{t}h$ and the $th$, $Zh$ channels, as well as $tZj$ and $Zt\bar{t}$ production at the 13 TeV LHC, and this conclusion is not sensitive to the possible new physics contribution induced by $Zt\bar{t}$ couplings in the $gg\to Zh$ production.
\end{abstract}

\maketitle

\noindent {\bf Introduction:~}
Verifying the electroweak symmetry breaking (EWSB) mechanism is one of the major tasks of particle physics at the Large Hadron Collider (LHC) after the discovery of the Higgs-like boson~\cite{Aad:2012tfa,Chatrchyan:2012ufa}.
In the Standard Model (SM), the EWSB is triggered by the Brout-Englert-Higgs mechanism, in which the couplings of the Higgs to EW gauge bosons play a crucial role. Although their coupling strengths are predicted by the SM, many new physics (NP) models could have a different prediction.  Observing a deviation in the gauge couplings from the SM prediction would shed light on various NP models and also the nature of EWSB.

Those gauge couplings are widely studied by both the theoretical~\cite{Englert:2014uua,Cheung:2014noa,Bergstrom:2014vla,Falkowski:2015fla,Craig:2015wwr,Corbett:2015ksa,Durieux:2017rsg,deBlas:2019wgy,Cao:2018cms} and experimental~\cite{Aad:2015gba,Khachatryan:2014jba,Aad:2019mbh,Sirunyan:2018koj,CMS-PAS-HIG-19-005} groups within global analysis of Higgs data under the $\kappa$-scheme or the SM effective field theory  (SMEFT) framework.  Recently, both the ATLAS~\cite{Aad:2019mbh}  and CMS~\cite{Sirunyan:2018koj, CMS-PAS-HIG-19-005} collaborations show a strong constraint for the gauge couplings through a combined analysis of Higgs production and decay signal strengths within $\kappa$-scheme at the 13 TeV LHC, i.e. $\kappa_W=1.10\pm 0.08,~\kappa_Z=1.05\pm 0.08$ (ATLAS) and $\kappa_W=1.10\pm 0.15,~\kappa_Z=0.99\pm 0.11$ (CMS) with an assumption $\kappa_{W,Z}>0$.
Here $\kappa_{W,Z}$ are gauge coupling strength modifiers of Higgs to the $W$ and $Z$ bosons, i.e.
\beq
\mathcal{L}_{hVV}=\kappa_W g_{hWW}^{\rm SM} hW^+_{\mu}W^{-\mu}+\frac{\kappa_Z}{2} g_{hZZ}^{\rm SM} hZ_{\mu}Z^{\mu},
\eeq 
where $g_{hVV}^{\rm SM}=2m_V^2/v$ with $V=W$, $Z$ being the gauge couplings in the SM and $v=246~{\rm GeV}$. The modifier $\kappa_{V}$ could be matched to the dimension-6 SMEFT operators after the EWSB~\cite{Buchmuller:1985jz,Giudice:2007fh,Grzadkowski:2010es}, and should be a leading approximation of the SMEFT to parametrize the new physics in Higgs gauge couplings~\cite{Arzt:1994gp}.
A global analysis to include Higgs, diboson and top quark measurements at the LHC in the framework of SMEFT with all possible dimension-6 operators could be found in Ref.~\cite{Ethier:2021bye}. With higher luminosity data being accumulated, one expects the accuracy on $\kappa_V$ could be further improved, e.g. the uncertainty will be reduced to 2\% at the high-luminosity LHC (HL-LHC)~\cite{ATLAS:2018jlh}, which operates at the $\sqrt{s}=14~{\rm TeV}$ with an integrated luminosity of $3~{\rm ab}^{-1}$. However, the analysis based on the current Higgs signal strengths and the simulation of the future colliders can only constrain the magnitude of $\kappa_V$, while not the relative sign between  $\kappa_W$ and $\kappa_Z$. It has been shown in Ref.~\cite{Low:2010jp} that a negative ratio $\lambda_{WZ}\equiv\kappa_W/\kappa_Z$ is also possible in the NP models. It is crucial to determine both the sign and the magnitude of $\kappa_{V}$ in order to further test the EWSB and search for the possible NP signals. 

\begin{figure}\centering
\includegraphics[scale=0.22]{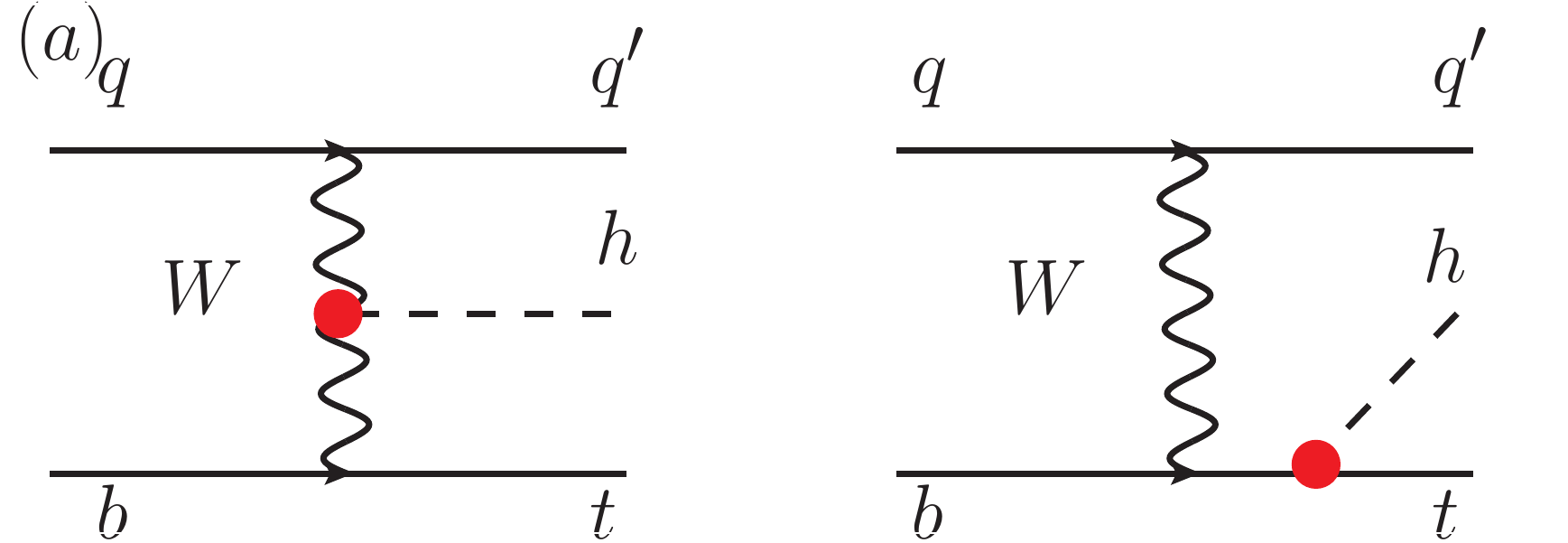}~
\includegraphics[scale=0.22]{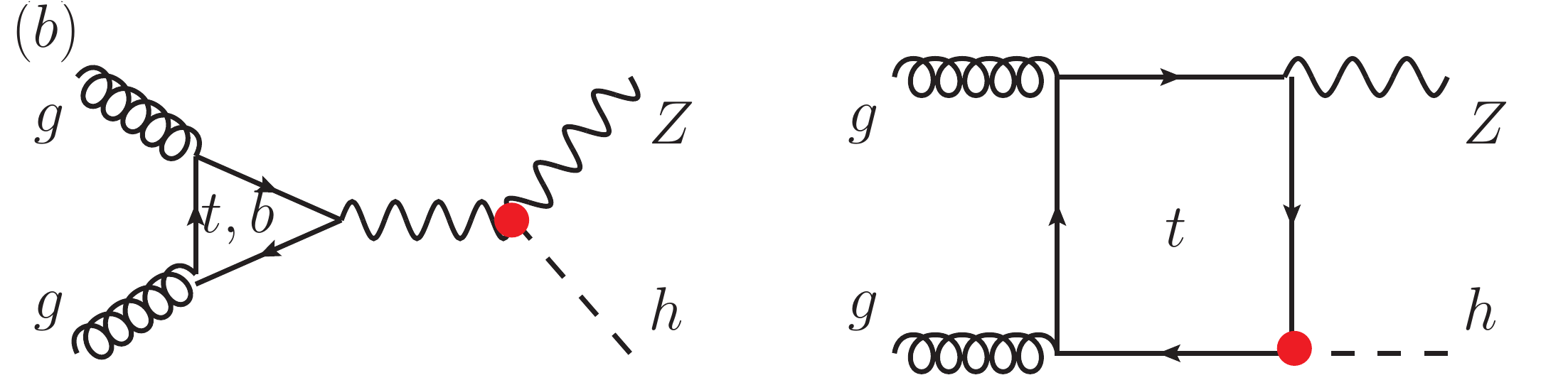}
\caption{Illustrative Feynman diagrams of $th$ (a) and $gg\to Zh$ (b) production at the LHC. The red dots denote the effective couplings including both the SM and NP effects. }
\label{fig:feyman}
\end{figure}

The sign of $\lambda_{WZ}$ could be resolved through the Higgs golden decay channel $h\to ZZ^*\to 4\ell$ with $\ell=e$, $\mu$, due to the interference effects between the tree and loop level processes~\cite{Chen:2016ofc}. Alternatively, one can also use $W^+W^-h$~\cite{Chiang:2018fqf} and vector bosons fusion production of $Vh$ processes~\cite{Stolarski:2020qim} at $e^+e^-$ colliders to determine the sign of $\lambda_{WZ}$. 
In this work, we propose  a novel method to pin down the sign of  $\lambda_{WZ}$ through the measurements of a Higgs boson with a single top quark ($th$) and $gg\to Zh$ production at the LHC; see Fig.~\ref{fig:feyman}. It is well known that the interference between the diagrams containing the $ht\bar t$ vertex and those containing the $hWW$ vertex in $th$ production is destructive when $\kappa_t$ and $\kappa_W$ have the same sign  due to the unitarity~\cite{Stirling:1992fx,Bordes:1992jy} (see Fig.~\ref{fig:feyman}(a)), where $\kappa_t$ is the modifier of top quark Yukawa coupling,
\beq
\mathcal{L}_{htt}=-\frac{m_t}{v}\kappa_th\bar{t}t.
\eeq
We can therefore measure the sign of the $ht\bar{t}$ coupling respect to that of the $hWW$ coupling through $th$ production at the LHC~\cite{Farina:2012xp,Kobakhidze:2014gqa,Rindani:2016scj,Degrande:2018fog,Barger:2018tqn,Barger:2019ccj,Maltoni:2019aot}.
Similarly the gluon-initiated $Zh$ production is sensitive to the relative sign between $ht\bar{t}$  and $hZZ$ couplings due to the cancelation between the box and triangle diagrams~\cite{Englert:2013vua,Hespel:2015zea,Goncalves:2015mfa,Harlander:2018yns,Vryonidou:2018eyv,Degrande:2018fog,Yan:2021veo}; see  Fig.~\ref{fig:feyman}(b). Therefore, it would be promising to probe the sign of $\lambda_{WZ}$ with the reference of $ht\bar{t}$ coupling through the measurements of $th$ and $gg\to Zh$ production at the LHC. 
We will demonstrate in the following that
combing the information of the $gg\to h$ production, $t\bar{t}h$ associated production and the two processes of we suggested, both the sign and magnitude of $\kappa_{V}$ could be well constrained.

\begin{figure}
\begin{center}
\includegraphics[scale=0.25]{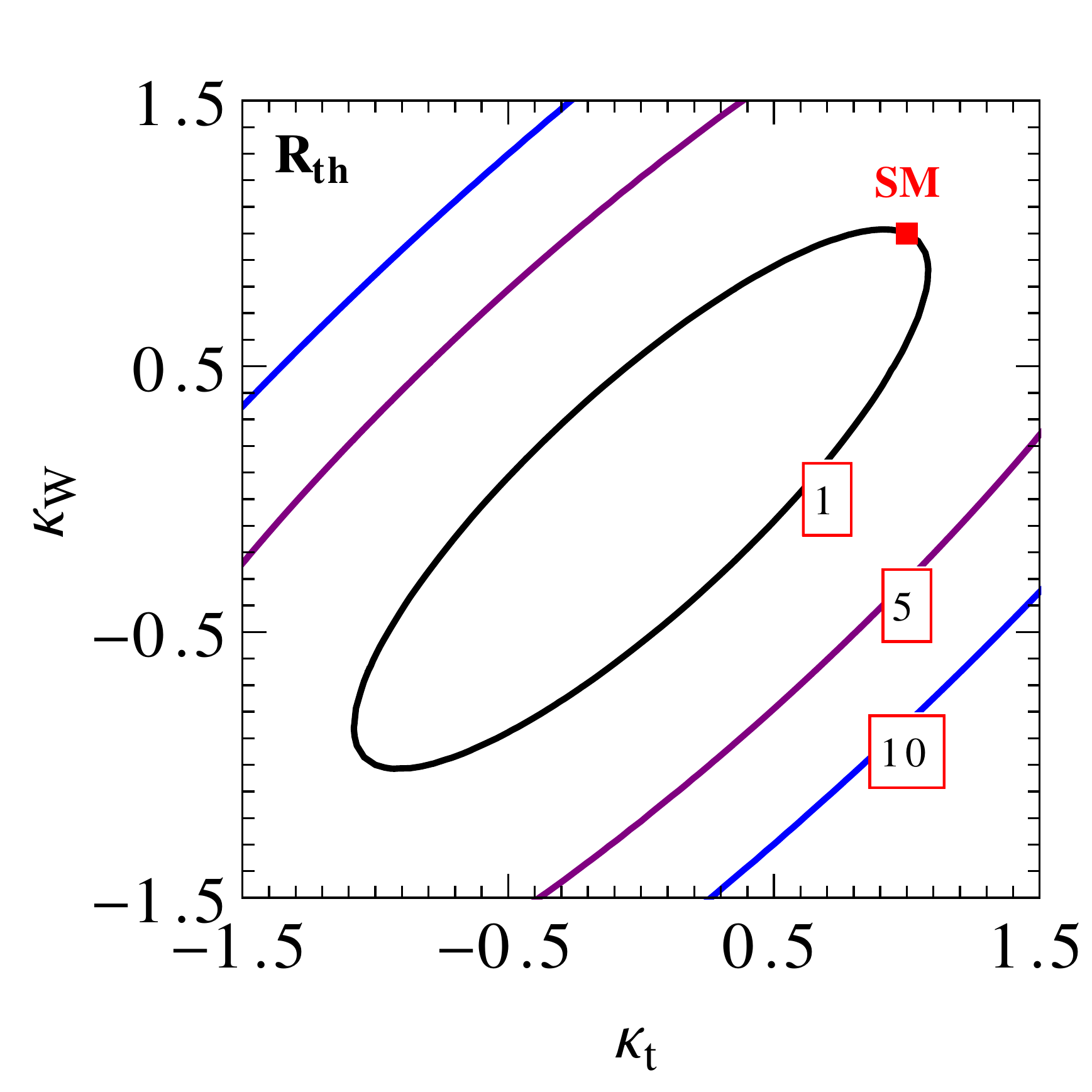}
\caption{The contours for $R_{th}=1,  5$ and 10 in the plane of anomalous couplings $\kappa_t$ and $\kappa_W$ at the 13 TeV LHC. }
\label{fig:Rth}
\end{center}
\end{figure}

\noindent {\bf $th$ production:~}
The $th$ associated production can be classified into three channels: $t$-channel, $s$-channel and $tW$-channel. The higher order QCD and EW corrections under the SM and SMEFT have been discussed in Refs.~\cite{Demartin:2015uha,Degrande:2018fog,Pagani:2020mov}. The three channels share the same subprocess of $bW^\mu\to th$ and are related to each other by crossing symmetry. At high energy limit, the amplitude of $bW^\mu\to h t$ scattering will be dominanted by the longitudinal polarized $W$ boson and  it could be written as,
\begin{multline}
M\sim \frac{1}{m_W^2}\bar{u}(t)\Big[m_t(\kappa_t-\kappa_W)\\
+\left(\frac{2m_W^2}{u}\kappa_W+\frac{m_t^2}{s}\kappa_t\right)\kslash_W\Big]P_L u(b).
\end{multline}
Here $s$, $t$, $u$ are the Mandelstam variables for describing the scattering of $bW\to th$. It clearly shows that there is a strong cancelation between $ht\bar{t}$ and $hWW$ anomalous couplings at high energy. As a result, the cross section of $th$ production can be significantly enhanced if the relative sign between $ht\bar{t}$ and $hWW$ is reversed. In order to compare $th$ cross section with non-standard $ht\bar{t}$ and $hWW$ couplings to the SM prediction, we define a ratio $R_{th}$ as,
\beq
R_{th}=\frac{\sigma(pp\to th)}{\sigma^{\rm SM}(pp\to th)}.
\eeq 
Note that we include all three channels in $R_{th}$ definition. Figure~\ref{fig:Rth} displays the contours of $R_{th}=1,5$ and 10 in the plane of anomalous couplings $\kappa_t$ and $\kappa_W$ with CT14LO PDF~\cite{Dulat:2015mca}. The $th$ production cross section could be enhanced up to one order of magnitude when  $\kappa_t\kappa_W<0$.

\begin{figure}
\begin{center}
\includegraphics[scale=0.24]{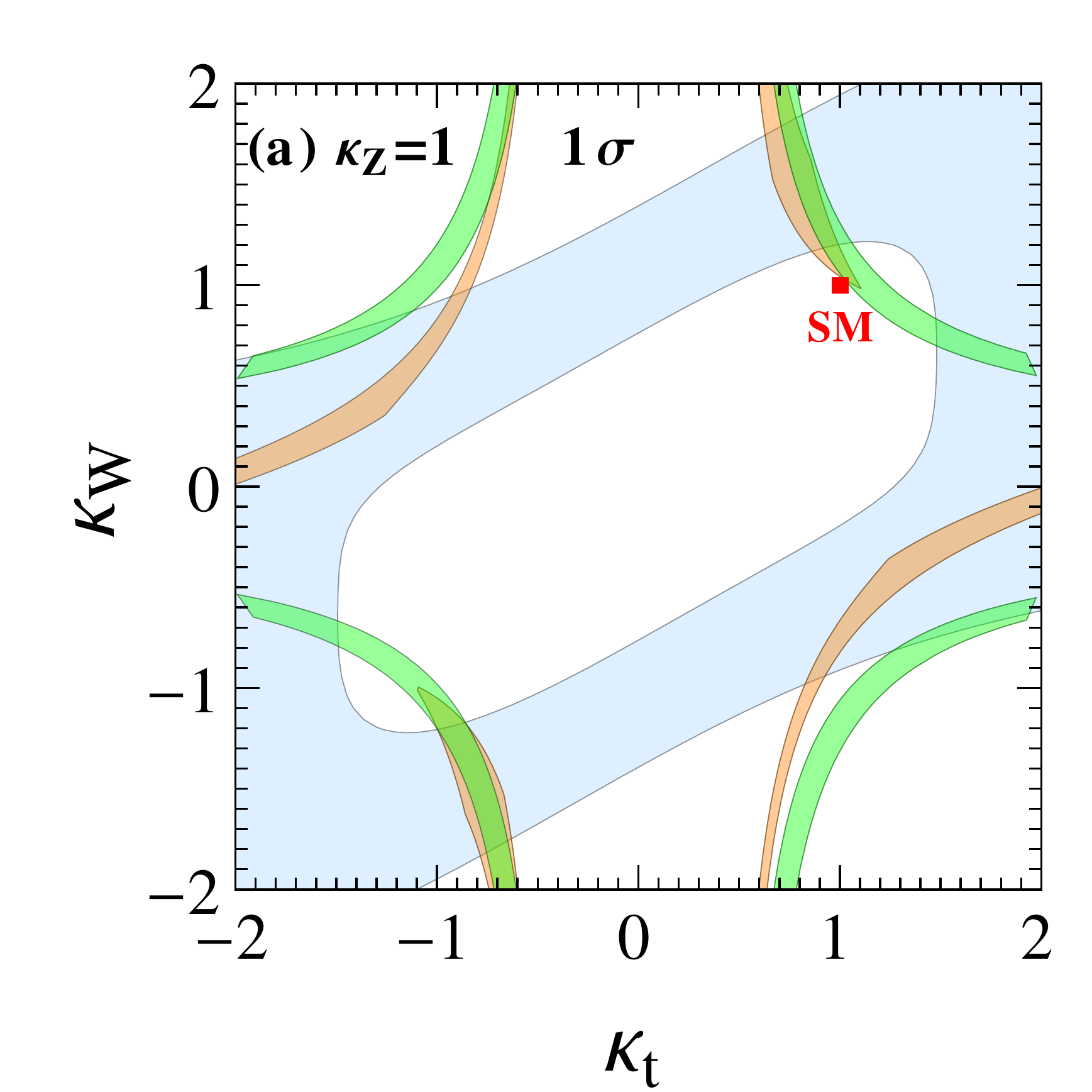}\,
\includegraphics[scale=0.24]{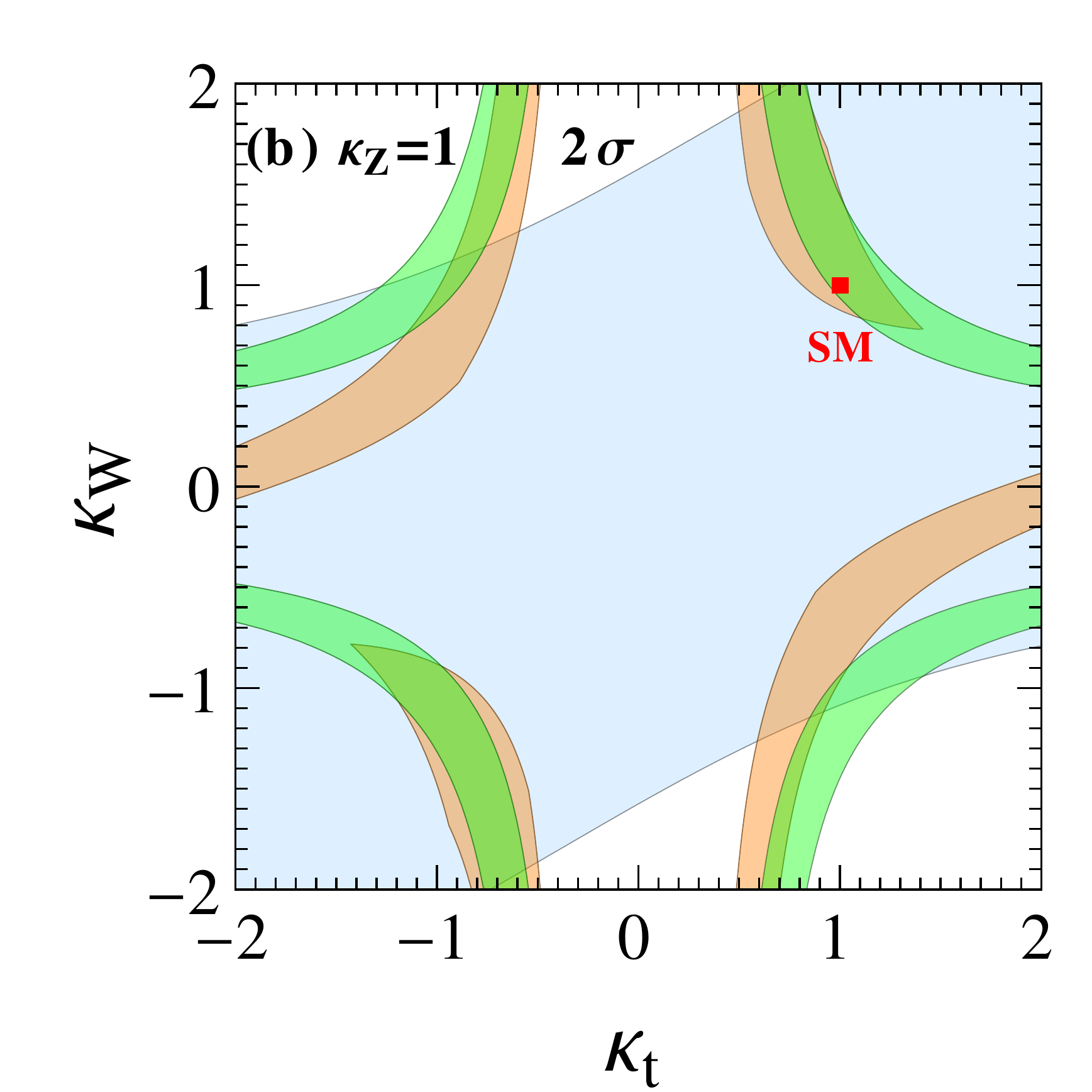}
\caption{Present constraints on the anomalous couplings $\kappa_t$ and $\kappa_W$ at the 13 TeV LHC. The light blue region comes from the $th$ cross section measurement~\cite{Sirunyan:2020icl}. The orange and green bands correspond to the limits from $t\bar{t}h$~\cite{Sirunyan:2020icl,Aad:2020ivc} and $gg\to h\to WW^*$~\cite{CMS-PAS-HIG-19-005}, respectively.}
\label{fig:thlimit}
\end{center}
\end{figure}

\begin{table*}
	\tabcolsep=5pt
	\begin{tabular}{c|c|c|c|c|c}
		\hline
		$th$~\cite{Sirunyan:2020icl} & $t\bar{t}h$~\cite{Sirunyan:2020icl} & $t\bar{t}h$~\cite{Aad:2020ivc} & ggF $(h\to WW^*)$~\cite{CMS-PAS-HIG-19-005} & ggF $(h\to ZZ^*)$~\cite{CMS-PAS-HIG-19-005} &-\\ \hline
		$5.7\pm 4.0$ & $0.92^{+0.26}_{-0.23}$ & $1.43^{+0.39}_{-0.34}$& $1.28^{+0.20}_{-0.19}$ & $0.98^{+0.12}_{-0.11}$ &-\\ \hline
		$Zh$~\cite{Aaboud:2018zhk}&$Zh$~\cite{Aad:2020jym}&$Zh$~\cite{Aad:2020eiv}&$Zh$~\cite{Aad:2020eiv}&$Zh$~\cite{Aaboud:2019nan}&$Zh$~\cite{Aaboud:2019nan}\\ \hline
		$0.92_{-0.26}^{+0.28}$&$1.08_{-0.23}^{+0.25}$&$0.34_{-0.70}^{+0.75}$&$0.28_{-0.83}^{+0.97}$&$1.6\pm 0.89$&$1.2\pm 0.34$\\ \hline
	\end{tabular}
	\caption{Signal strengths of Higgs production at the 13 TeV LHC.}
	\label{tbl:sig}
\end{table*}

Recently, the $th$ signal strength has been measured at the 13 TeV LHC by both the CMS ($137~{\rm fb}^{-1}$)~\cite{Sirunyan:2020icl} and ATLAS ($139~{\rm fb}^{-1}$)~\cite{Aad:2020ivc} collaborations, and the most stringent limit comes from the former, which is $\mu(th)=5.7\pm 2.7 ~({\rm stat})\pm3.0 ~({\rm syst})$. In Fig.~\ref{fig:thlimit}, we compare the precision on the determination of the Higgs anomalous couplings $\kappa_t$ and $\kappa_W$ via the measurements of inclusive cross section from $th$~\cite{Sirunyan:2020icl} (light blue), $t\bar{t}h$~\cite{Sirunyan:2020icl,Aad:2020ivc} (orange) and $gg\to h\to WW^*$~\cite{CMS-PAS-HIG-19-005} (green), assuming $\kappa_Z=1$. We summarize the signal strengths of Higgs production at the 13 TeV LHC in table~\ref{tbl:sig}.
The higher order QCD correction for $th$ production processes have been included by a constant $k$-factor. 
A detail analysis of QCD correction for each anomalous couplings can be found in Ref.~\cite{Degrande:2018fog} and it shows that a constant $k$-factor should be a good approxiamtion to parametrize the QCD effects. Furthermore, the scale and PDF uncertainties are around few percentage level at the NLO accuracy~\cite{Degrande:2018fog}, and the results from 4-flavor and 5-flavor scheme provide fully consistent and similarly precise predictions for the total cross section and distributions~\cite{Demartin:2015uha}. Therefore, we expect the conclusion in this section should not strongly depend on those theoretical uncertainties.
From Fig.~\ref{fig:thlimit}, it is evident that the current measurements have favored same-sign $\kappa_t$ and $\kappa_W$ at around $2\sigma$ level, i.e. $\kappa_t\kappa_W>0$ is required. 

We remark that though we assume $\kappa_Z=1$ in the analysis, the sign of $\kappa_t\kappa_W$ should not strongly dependent on this assumption since $\kappa_Z$ will only change the Higgs total decay width, while not for the $th$ scattering cross section. Moreover, the magnitude of $\kappa_Z$ has been constrained severely at the LHC~\cite{Aad:2019mbh,Sirunyan:2018koj, CMS-PAS-HIG-19-005}.

\noindent{\bf  $Zh$ production via gluon fusion:~} 
We consider the $ht\bar{t}$, $hZZ$ and $Zt\bar{t}$ couplings to the $gg\to Zh$ production. The couplings of top quark to $Z$ boson could be parametrized generically with,
\beq
\mathcal{L}_{Ztt}=\frac{g_W}{2c_W}\bar{t}\gamma_\mu(\kappa_v^tv_t-\kappa_a^ta_t\gamma_5)tZ_\mu,
\eeq
where $g_W$ is the EW gauge coupling and $c_W$ is the cosine of the weak mixing angle $\theta_W$. The vector and axial-vector couplings of $Z$ boson to top quark in the SM are $v_t=1/2-4/3s_W^2$ and $a_t=1/2$.
The helicity amplitudes of $g(\lambda_1)g(\lambda_2)\to Z(\lambda_3)h$ with helicity $\lambda_i=\pm,0$ for particle $i$ have been calculated in Refs.~\cite{Kniehl:1990iva,Kniehl:2011aa,Yan:2021veo}. It shows that the dominant amplitudes come from $(\pm,\pm,0)$ helicity configurations and the results with $m_b=0$ are~\cite{Yan:2021veo},
\begin{align}
M_{++0}^{\triangle}&=2\frac{\sqrt{\lambda}}{m_Z}\sum_{t,b}\left[\kappa_a^q\kappa_Z\frac{a_{q}g_{hZZ}^{\rm SM}}{m_Z^2}\left(F_\triangle(s,m_q^2)+2\right)\right]N,\nn\\
M_{++0}^{\square} &=\frac{4m_t^2}{m_Zv\sqrt{\lambda}}\kappa_a^t\kappa_ta_t\left[F_{++}^0+(t\leftrightarrow u)\right]N,
\label{eq:hel}
\end{align}
where
\bea
\lambda&=&s^2+m_Z^4+m_h^4-2(sm_Z^2+m_Z^2m_h^2+m_h^2s),\nn\\
N&=&\frac{\alpha_s g_W}{32\pi c_W}.
\eea 
The symbols $\triangle$ and $\square$ denote the contributions from triangle and box diagrams, respectively (see Fig.~\ref{fig:feyman}(b)). 
The parameter $a_b=-1/2$ is the axial-vector coupling of $Z$ boson to bottom quark and parameter $\kappa_a^b=1$.
Note that the helicity amplitudes $M_{--0}^{\triangle,\square}$ could be related to $M_{++0}^{\triangle,\square}$ by Bose symmetry~\cite{Kniehl:2011aa}. 
The definition of the scalar functions $F_\triangle$ and $F_{++}^0$ in Eq.~(\ref{eq:hel}) could be found in Ref.~\cite{Kniehl:2011aa}.
We should note  that only the axial-vector component ($\kappa_a^t$) of the $Zt\bar{t}$ couplings can contribute to the $gg\to Zh$ production due to the charge conjugation invariance~\cite{Yan:2021veo}. 

At high energy limit, only the top quark contributes to the $gg\to Zh$ scattering and the total amplitude is,
\beq
M_{\pm,\pm,0}\sim\frac{m_t^2}{m_Z^2}\left(\kappa_Z-\kappa_t\right)\log^2\left(-\frac{s}{m_t^2}\right),
\eeq
hence a strong cancellation occurs between the triangle and box diagrams in the SM where $\kappa_t=\kappa_Z=1$. However, such relation could be violated in the NP models, so that the cancelation is spoiled and the $Zh$ cross section would be enhanced.

\begin{figure}
\begin{center}
\includegraphics[scale=0.25]{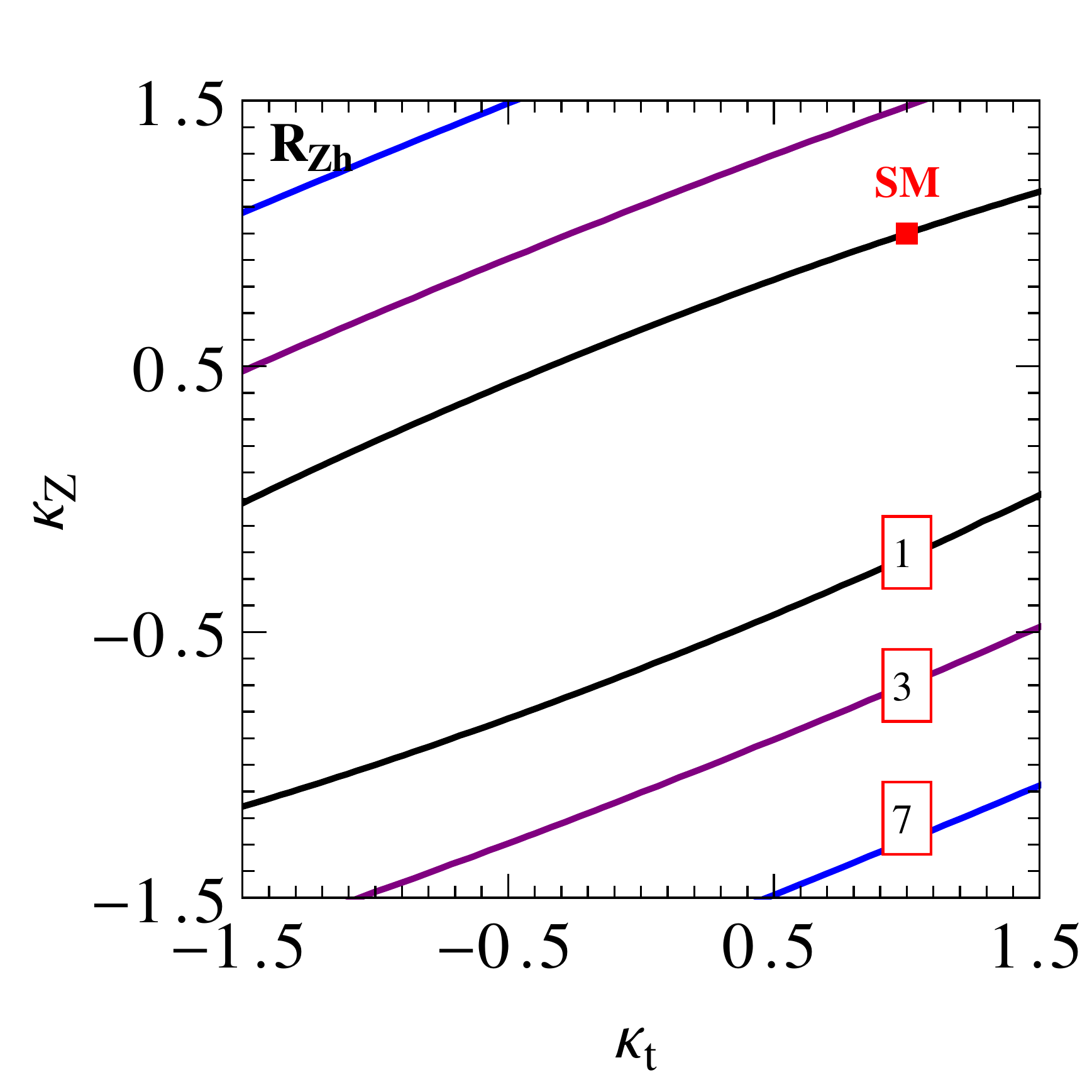}
\caption{The contours for $R_{Zh}=1,  3$ and 7 in the plane of anomalous couplings $\kappa_t$ and $\kappa_Z$ at the 13 TeV LHC. }
\label{fig:Rzh}
\end{center}
\end{figure}

Similar to $R_{th}$, we define a ratio $R_{Zh}$ to compare the $Zh$ scattering cross section with the SM prediction,
\beq
R_{Zh}=\frac{\sigma(gg\to Zh)}{\sigma^{\rm SM}(gg\to Zh)}.
\eeq
Figure~\ref{fig:Rzh} displays the contours of $R_{Zh}=1,3$ and 7 with $\kappa_a^t=1$ and CT14LO PDF~\cite{Dulat:2015mca} in the plane of anomalous couplings $\kappa_t$ and $\kappa_Z$.  It shows that the cross section could be enhanced about few times compared to the SM prediction in the parameter space $\kappa_t\kappa_Z<0$.
On the other hand the $gg\to Zh$ production contributes $\sim15\%$ to the total cross section of the $pp\to Zh$ process in the SM at the 13 TeV LHC. Therefore, few times enhancement of $gg\to Zh$ is a large enough deviation that can be detected at the LHC.

\begin{figure}
\begin{center}
\includegraphics[scale=0.24]{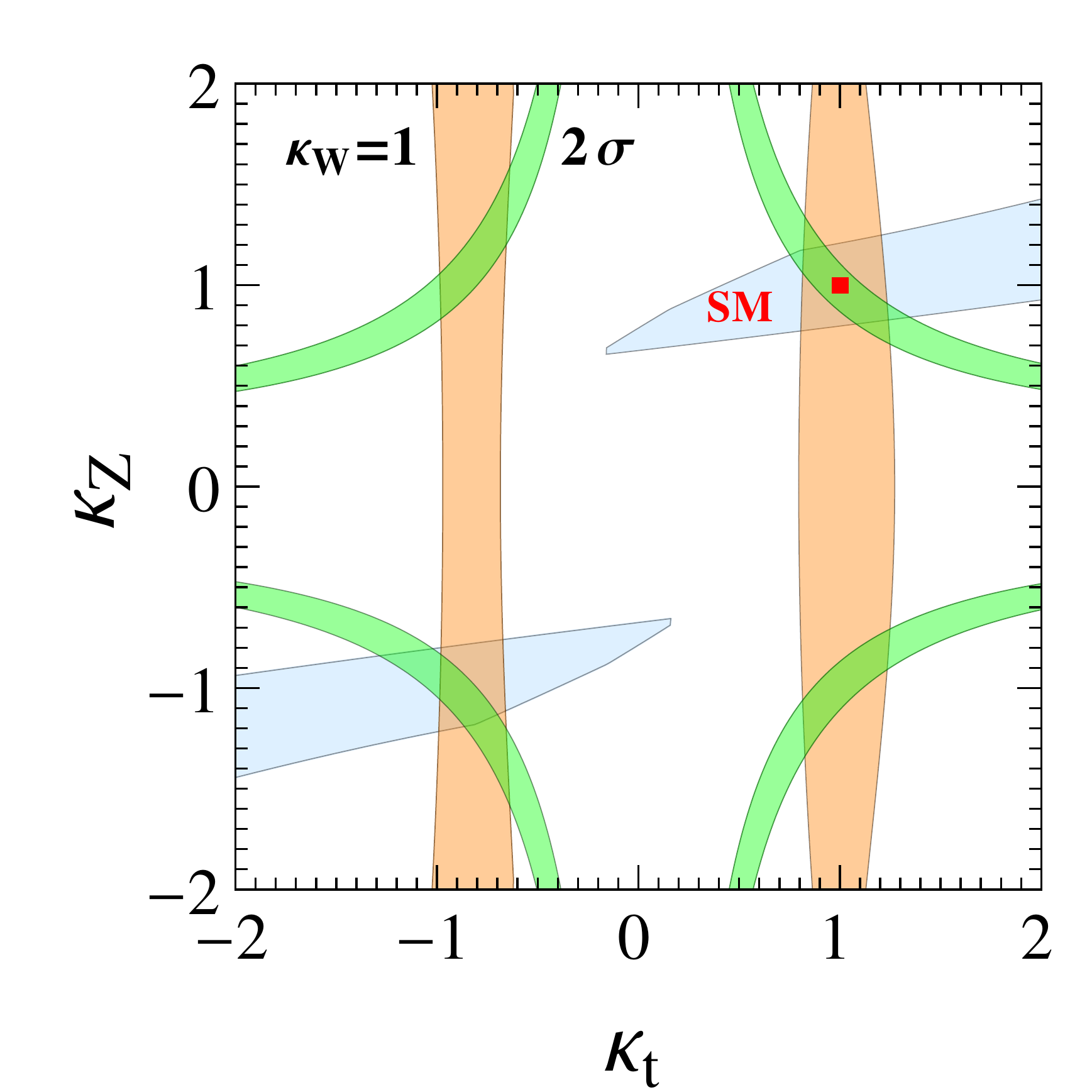}
\caption{Present constraints on the anomalous couplings $\kappa_t$ and $\kappa_Z$ with $\kappa_a^t=1$ at the 13 TeV LHC. The light blue region comes from the inclusive cross section and transverse momentum  distribution of $Z$ boson in the $pp\to Zh$ production~\cite{Aaboud:2018zhk,Aaboud:2019nan,Aad:2020eiv,Aad:2020jym,CMS-PAS-HIG-19-005}. The orange and green bands are corresponding to the limits from $t\bar{t}h$~\cite{Sirunyan:2020icl,Aad:2020ivc} and $gg\to h\to ZZ^*$~\cite{CMS-PAS-HIG-19-005} production, respectively.}
\label{fig:zhlimit}
\end{center}
\end{figure}

We note that both the inclusive cross section and transverse momentum distribution of $Z$ boson in the $pp\to Zh$ production at the 13 TeV LHC have been measured by the ATLAS and CMS collaborations with integrated luminosities $79.8\sim139~{\rm fb}^{-1}$~\cite{Aaboud:2018zhk,Aaboud:2019nan,Aad:2020eiv,Aad:2020jym,CMS-PAS-HIG-19-005}. We show the limits from the present measurements to the plane of anomalous couplings $\kappa_t$ and $\kappa_Z$ with assumption $\kappa_W=\kappa_a^t=1$ at $2\sigma$ level in Fig.~\ref{fig:zhlimit}. The light blue region denotes the constraint from the measurements of the $pp\to Zh$ production, in which both $q\bar q$ and $gg$ initial states are considered. A constant $k$-factor has been used to mimic the higher order QCD correction effects for both $q\bar{q}\to Zh$ and $gg\to Zh$ production in the analysis, i.e. $k_{qq}=1.3$ and $k_{gg}=2.7$~\cite{deFlorian:2016spz,Hasselhuhn:2016rqt}. 
It is worthwhile discussing how much our result will be influenced by the QCD corrections. The NNLO QCD corrections to the $Zh$ production with the anomalous couplings have been discussed in Ref.~\cite{Bizon:2021rww} and it shows a constant $k$-factor should be a reasonable assumption in this work~\cite{Bizon:2021rww}. Furthermore, the scale uncertainty is around $1\%\sim 2\%$, as a result, the high order QCD effects should not alter the conclusion in this section. The orange and green bounds show the constraints imposed by the measurements of $t\bar{t}h$~\cite{Sirunyan:2020icl,Aad:2020ivc} and $gg\to h\to ZZ^*$ production~\cite{CMS-PAS-HIG-19-005}; (see  tabel~\ref{tbl:sig} for the detail of the signal strengths.) It clearly shows that the current measurements of the $Zh$ cross sections at the LHC has resolved  the ambiguity of the relative sign between $\kappa_t$ and $\kappa_Z$, i.e. $\kappa_t\kappa_Z>0$ is allowed. 
Again, we emphasize that the sign $\kappa_t\kappa_Z$ should not be sensitive to the assumption of $\kappa_W=1$ due to $\kappa_W$ can not change the cross section of $Zh$ scattering.

\begin{figure}
\begin{center}
\includegraphics[scale=0.24]{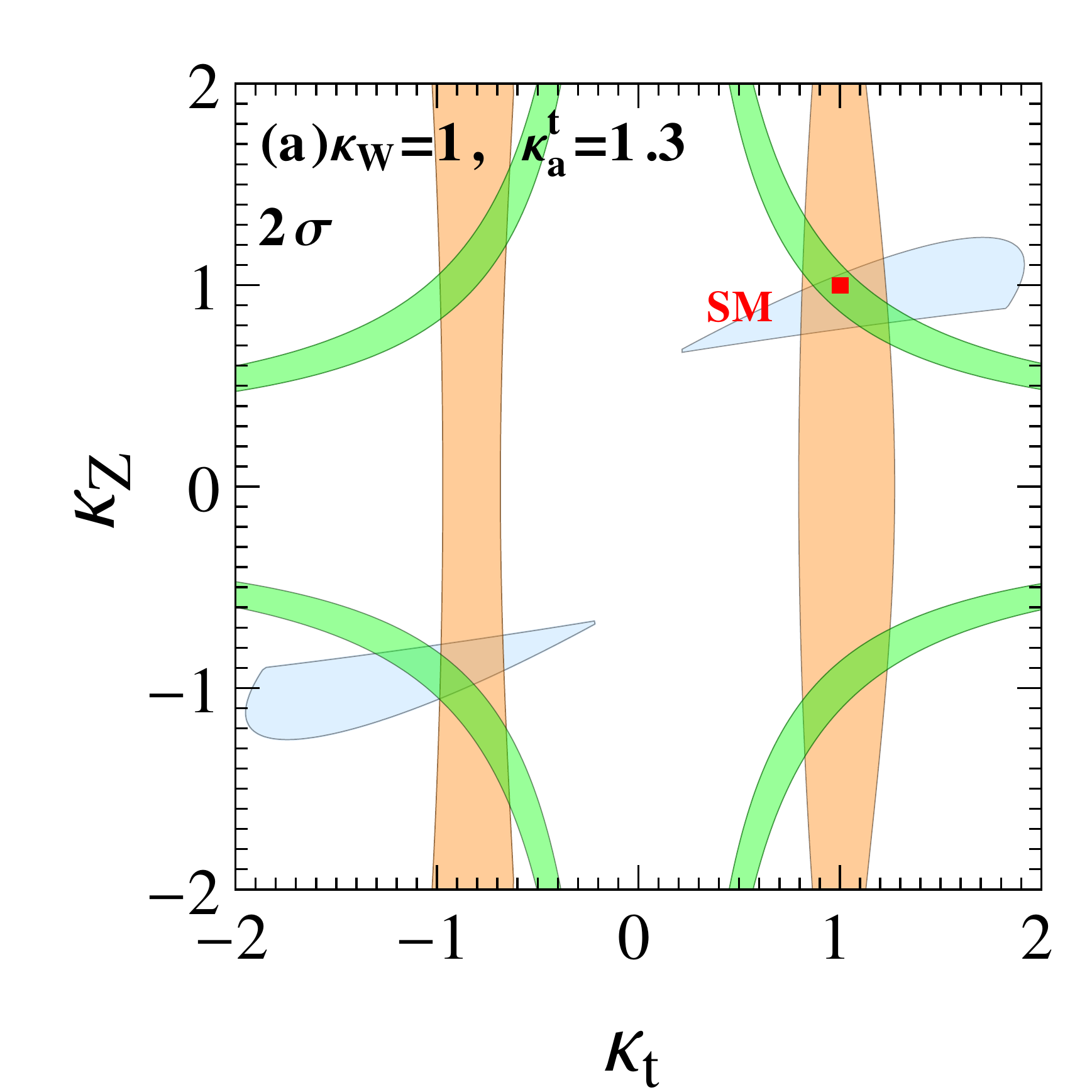}\,
\includegraphics[scale=0.24]{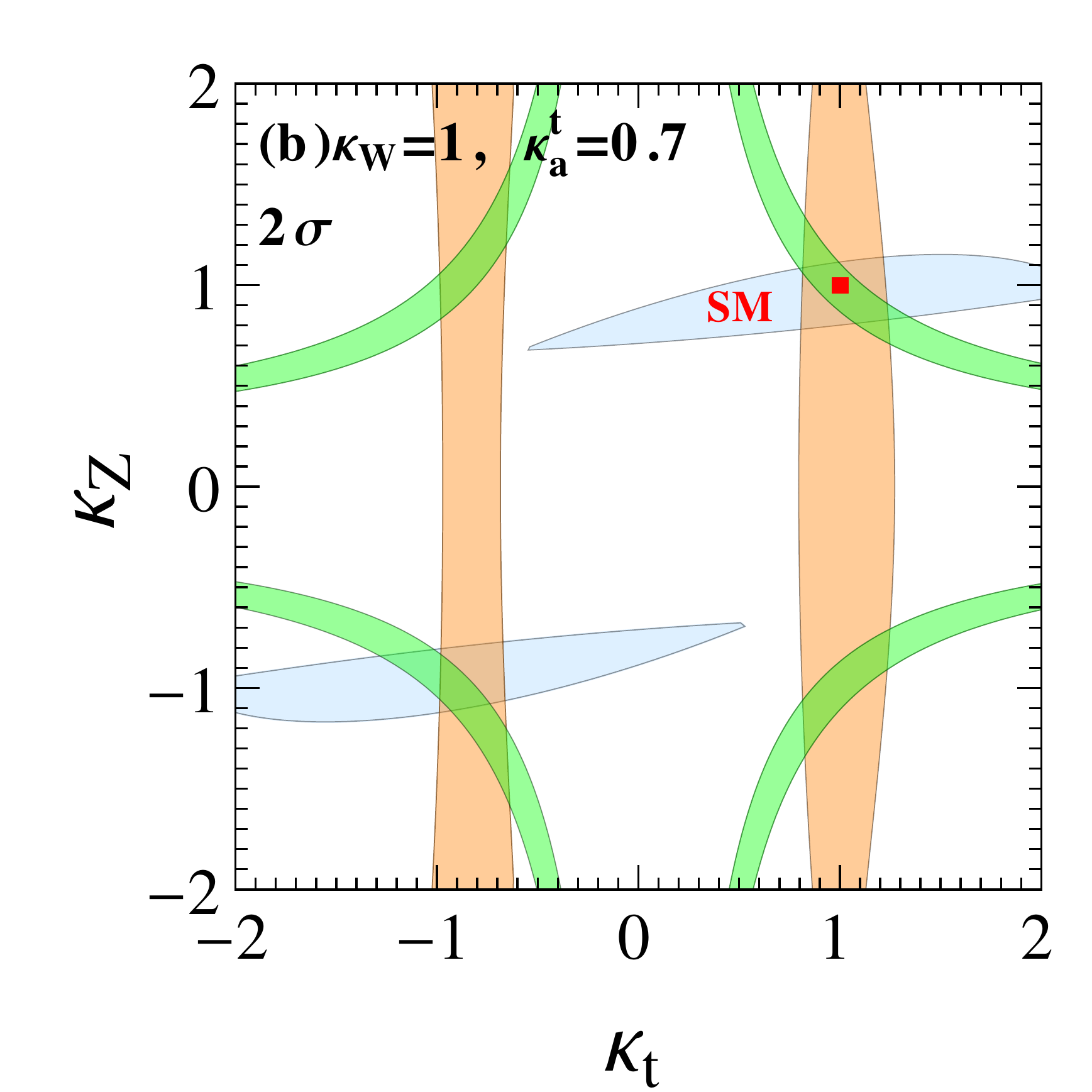}
\caption{Similar to Fig.~\ref{fig:zhlimit}, but for $\kappa_a^t=1.3,0.7$.}
\label{fig:zhlimit2}
\end{center}
\end{figure}

Next we consider the impact of the non-standard $Zt\bar{t}$ coupling to determine the relative sign between $\kappa_t$ and $\kappa_Z$.
The $Zt\bar{t}$ couplings have been well constrained by the measurements of $tZj$~\cite{Aad:2020wog,Sirunyan:2018zgs} and $Zt\bar{t}$~\cite{CMS:2019too,Aaboud:2019njj} productions at the 13 TeV LHC. The limits could be potentially improved after we combining the measurement from $gg\to ZZ$ production~\cite{Cao:2020npb}.  As a conservative estimation of the impact from the $Zt\bar{t}$ coupling, we choose  two benchmark points of $\kappa_a^t=0.7,1.3$ in the analysis, and show the allowed parameter space of $\kappa_t$ and $\kappa_Z$ at $2\sigma$ level with above value of $\kappa_a^t$ in Fig.~\ref{fig:zhlimit2}.
Although the value of $\kappa_a^t$ will change the allowed parameter space of the $\kappa_t$ and $\kappa_Z$ from the $Zh$ measurements, the relative sign between them is still fixed, i.e. $\kappa_t\kappa_Z>0$.

\begin{figure*}
	\begin{center}
		\includegraphics[scale=0.4]{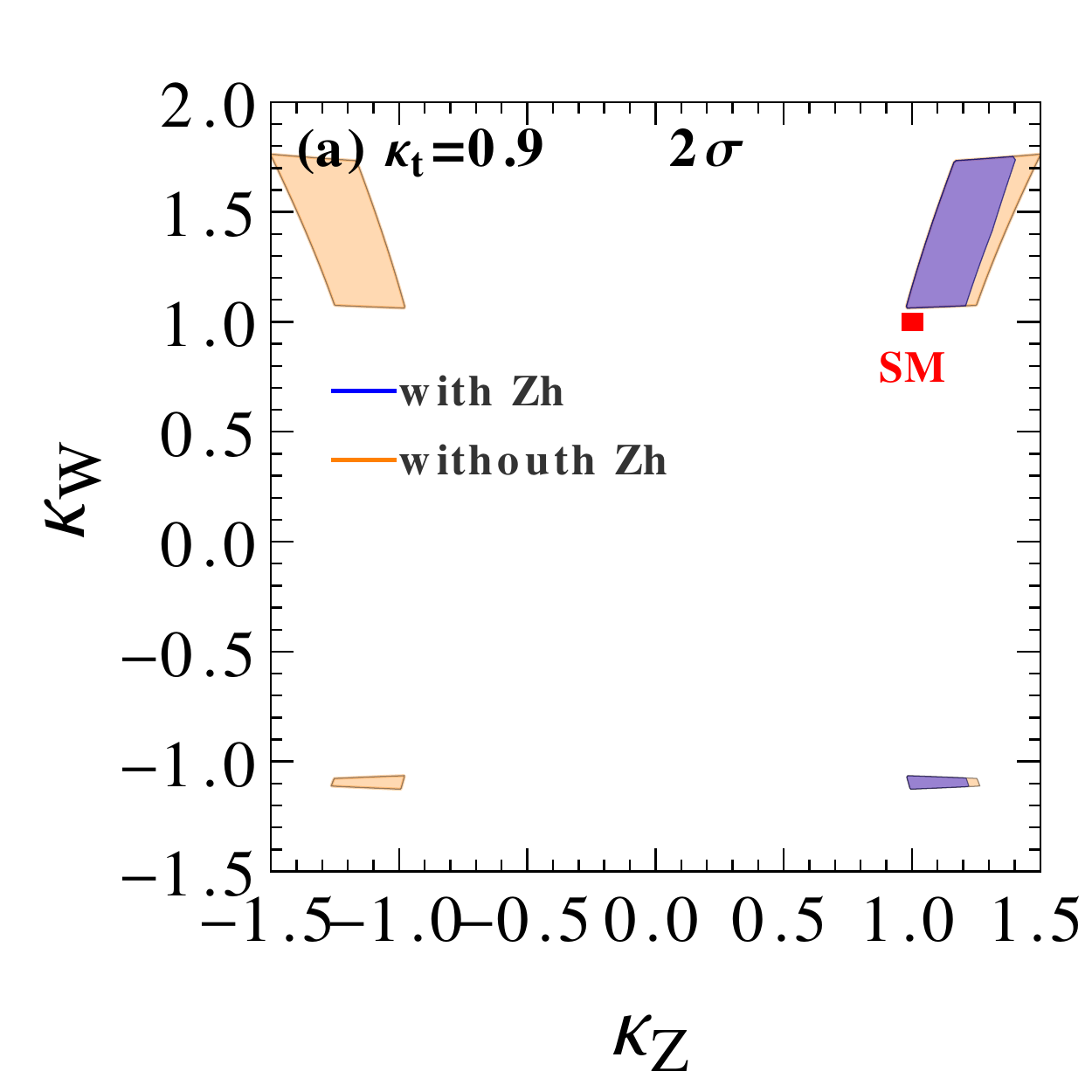}
		\includegraphics[scale=0.4]{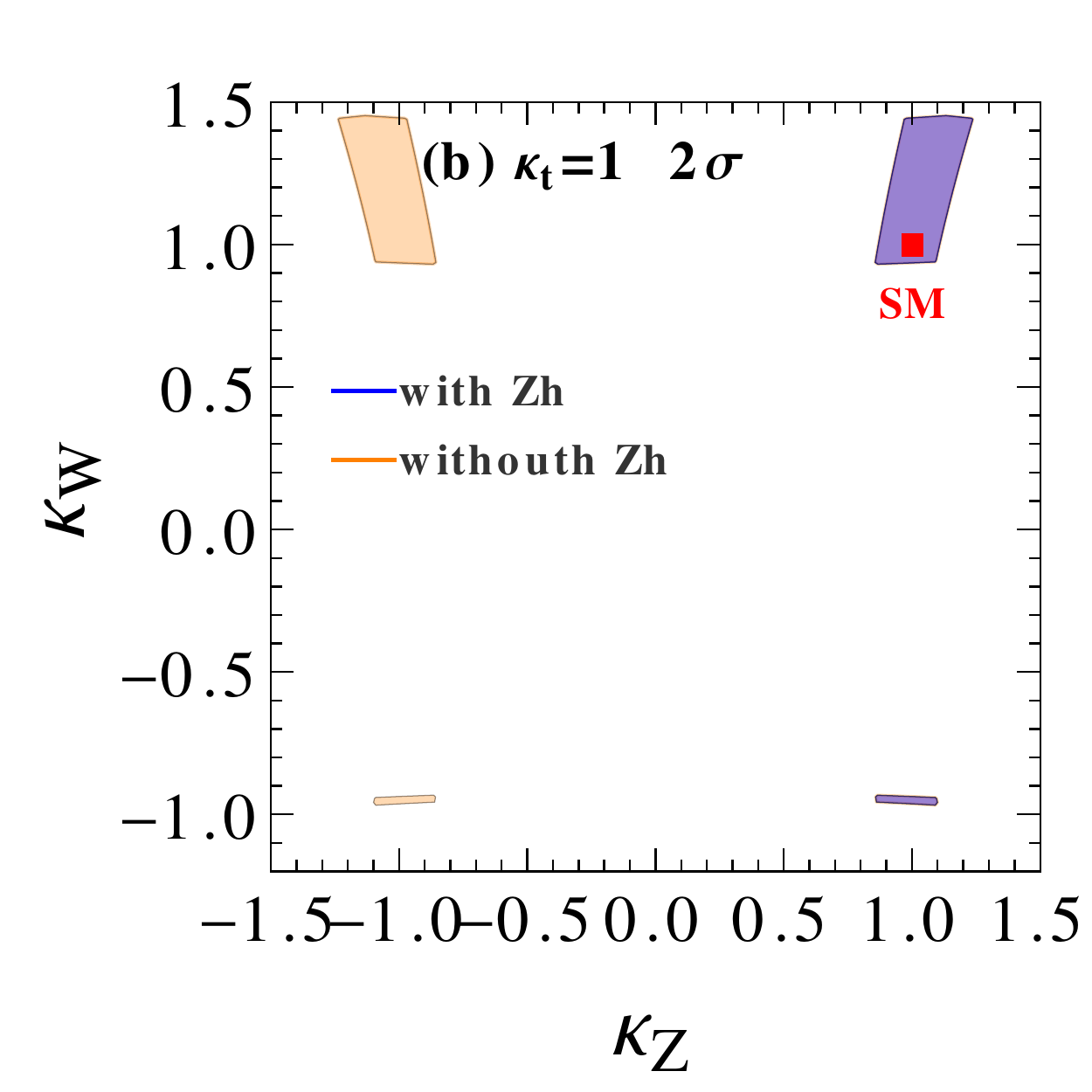}
		\includegraphics[scale=0.4]{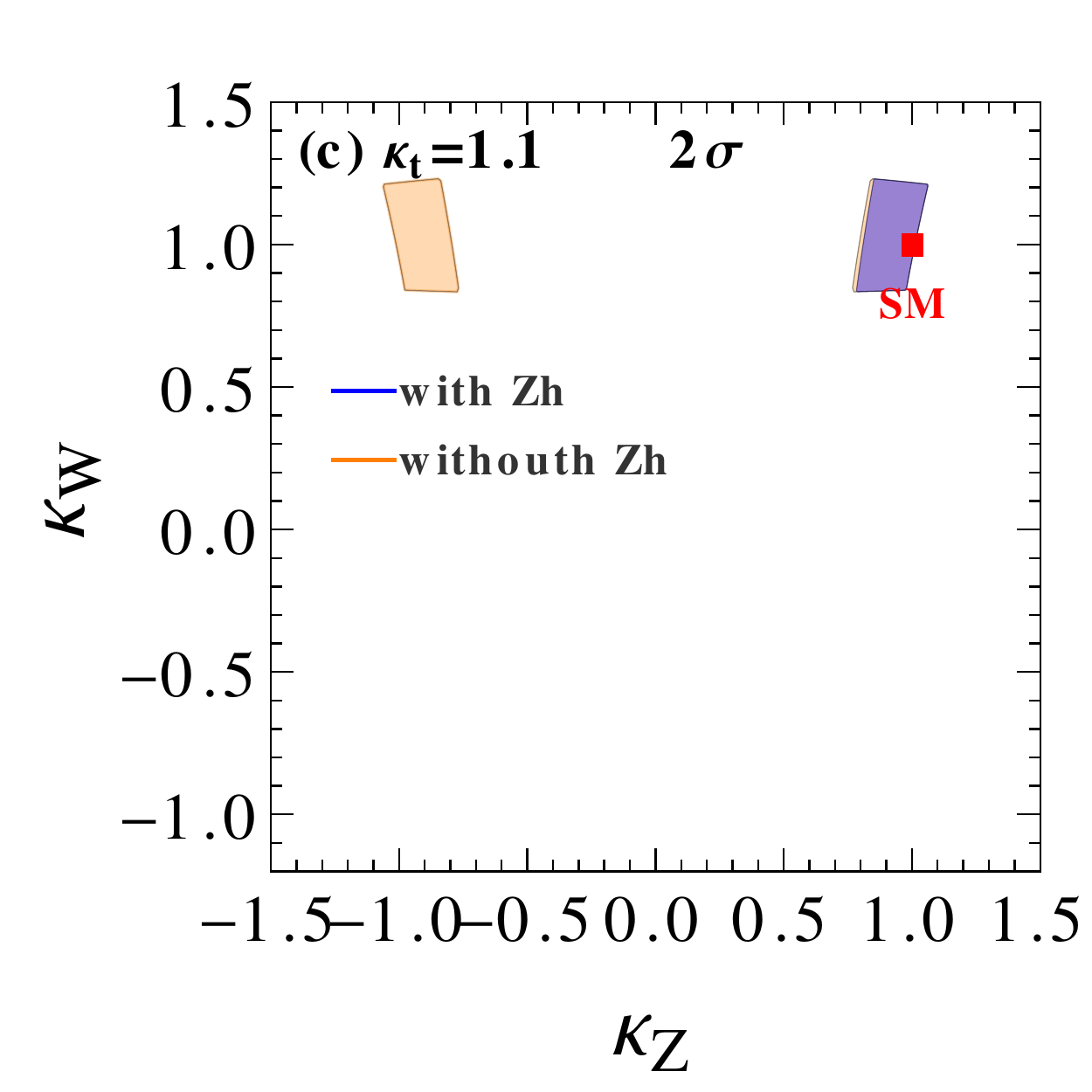}
		\caption{Present constraints on the anomalous couplings $\kappa_Z$ and $\kappa_W$ at the 13 TeV LHC with $\kappa_t=0.9,1,1.1$ and $\kappa_a^t=1$. 
			The blue region comes from the limits after we include all the data, while the orange band denotes the impact after we removing the $Zh$ data (both the inclusive cross section and transverse momentum distribution of $Z$ boson in $pp\to Zh$ production).}
		\label{fig:combine}
	\end{center}
\end{figure*} 

\noindent {\bf Summary and discussion:~}
Now equipped with the constraints for the Higgs couplings $ht\bar{t}$ and $hWW$ (see Fig.~\ref{fig:thlimit}), $ht\bar{t}$ and $hZZ$ (see Fig.~\ref{fig:zhlimit}) at the 13 TeV LHC, we are ready to estimate the potential of pining down the sign of $\lambda_{WZ}$ through the global analysis of the $gg\to h$, $t\bar{t}h$ production and $th$, $Zh$ scattering with present measurements.  From the above discussion one sees that current data favors same sign for  both the ($ht\bar{t}$, $hWW$) and ($ht\bar{t}$, $hZZ$) couplings,  as a result, the $ht\bar{t}$ coupling could be a good reference to determine the relative sign between Higgs gauge couplings. In Fig.~\ref{fig:combine}, we show the constraints on the plane of $\kappa_Z$ and $\kappa_W$ with $\kappa_t=0.9,1,1.1$ and $\kappa_a^t=1$ from the current measurements with (blue) and without $Zh$ data (orange) at  $2\sigma$ level. Although the $Zh$ data itself can not improve the accuracy of the $\kappa_V$, the $\lambda_{WZ}<0$ region could be excluded almost at $2\sigma$ level by $Zh$ measurements, and this conclusion is not sensitive to possible new physics contribution induced by $Zt\bar{t}$ coupling in the $gg\to Zh$ production (see Fig.~\ref{fig:zhlimit2}). At the HL-LHC, all the experimental measurements could be much improved compared to the current data, and  as a result, we  expect that the gauge couplings of Higgs to $W$ and $Z$ bosons could be well constrained and the nature of EWSB will surface at that time. 

\noindent{\bf Acknowledgements:~}
The authors thank Yandong Liu for helpful discussions and comments. B.~Y. is supported by the U.S. Department of Energy, Office of Science, Office of Nuclear Physics, under Contract DE-AC52-06NA25396, [under an Early Career Research Award (C. Lee),] and through the LANL/LDRD Program. K.~P.~X. is supported by the Grant Korea NRF-2019R1C1C1010050.

\bibliographystyle{apsrev}
\bibliography{reference}

\end{document}